# Rational design of metallic nanocavities for resonantly enhanced four-wave mixing

Euclides Almeida and Yehiam Prior

*Department of Chemical Physics, Weizmann Institute of Science, Rehovot 76100, Israel*

*Euclides.almeida@weizmann.ac.il ; Yehiam.prior@weizmann.ac.il*

**ABSTRACT**

Optimizing the shape of nanostructures and nano-antennas for specific optical properties has evolved to be a very fruitful activity. With modern fabrication tools a large variety of possibilities is available for shaping both nanoparticles and nanocavities; in particular nanocavities in thin metal films have emerged as attractive candidates for new metamaterials and strong linear and nonlinear optical systems. Here we rationally design metallic nanocavities to boost their Four Wave Mixing response by resonating the optical plasmonic resonances with the incoming and generated beams. The linear and nonlinear optical responses as well as the propagation of the electric fields inside the cavities are derived from the solution of Maxwell's equations by using the 3D finite-differences time domain method. The observed conversion-efficiency of near infra-red to visible light equals or surpasses that of BBO of equivalent thickness. Implications to further optimization for efficient and broadband ultrathin nonlinear optical materials are discussed.

Keywords: Plasmonics, Four Wave Mixing, Nanostructures, Nonlinear FDTD calculations

## INTRODUCTION

The rational design of optical metamaterials with dimensions smaller than or comparable to the wavelength of light is fundamental for future applications that require miniaturization of photonic devices[1, 2] [3]. While the knowledge of the linear optical properties of these materials has advanced tremendously in the last decade, the nonlinear optical properties are known to a lesser extent. The strong intrinsic nonlinear response of noble metals makes them good candidates for nonlinear optical applications at the nanoscale[4]. The electric field enhancement at plasmon resonance frequencies can be used to boost the already strong nonlinear response of the noble metals by many orders of magnitude [5] [6]. Furthermore, resonant metallic nanostructures can be combined with other nonlinear materials for enhanced nonlinear generation [7, 8] , opening up a myriad of possibilities for the engineering of very small, efficient and broadband nonlinear materials.

Second harmonic generation (SHG) has been observed in several metallic nanostructures like arrays of nanoparticles[9] and nanoholes[5, 10], gold "nanocups" [11] and single nanocavities[12, 13] of different shapes. As SHG requires lack of spatial symmetry, the signal in these nanostructures is generated only at the interfaces, where the symmetry is broken, decreasing the overall efficiency of SHG. On the other hand, higher odd-order nonlinear phenomena, like third-harmonic generation and four-wave mixing (FWM), can occur in non-centrosymmetric media, and has been proposed for high-sensitivity nonlinear spectroscopy[14].

There are different strategies to enhance the optical nonlinear generation in nanostructures and metallic surfaces [6], [15-18]. As in any nonlinear process, the signal strongly depends on the intensity of the input beams, and therefore the material must be tailored to obtain a large field enhancement at the fundamental frequencies. Equally important is the optical response of the material at the frequency of the nonlinear signal, which must not absorb or attenuate the signal[10]. In the case of coherent NL processes, like SHG and FWM, phase matching of the involved fields is required, thus the metamaterial must be designed to reduce phase mismatch[18].

Here we show that efficient FWM generation can be obtained in an array of metallic nanorectangles. Control of the nanocavities geometrical parameters enables the tuning of the optical resonances to the frequencies of the interacting fields to obtain a stronger FWM signal. Extraordinary optical transmission[19] (EOT) contributes by increasing the local fields inside the nanoholes and facilitating the propagation of the FWM signal. For demonstration purposes, we limited this study to a single geometrical shape (rectangles), kept the fraction of milled area constant and we only tried to optimize a single parameter – the aspect ratio (AR) of the rectangles. Clearly, the parameter space to be optimized can be much larger (shape, area, film thickness, etc.) but as an illustrative example of rational design we preferred to limit the optimization to a single parameter. The design goes through a numerical calculation of the dependence of the FWM signal on the AR using a nonlinear 3D finite differences time-domain (NL-3D-FDTD) model. Following the design, the fabricated optimized shapes are measured experimentally, and the agreement is found to be very good. Finally, an intuitive mode theory which takes into account field distribution and phase matching[10] is implemented to our case of FWM in nanocavities. The model provides insight into the observed results and suggests strategies for further optimization of efficient nonlinear metamaterials.

## RESULTS

Periodic square arrays of 45x45 rectangular holes were milled in a high quality 250 nm thick free-standing gold film using focused ion beam (FIB). The periodicity of the holes within an array was set to 510 nm. For each array, the aspect ratio (AR) of the rectangles is different, but the open area is kept constant and equal to that of square holes of 200nm x 200nm (AR=1). Several configurations were calculated and the corresponding experiments were performed. For each array, the linear transmission spectrum was predicted and measured for different input light polarization. The FWM at $\omega_{FWM} = 2\omega_1 - \omega_2$ was measured for two different input frequency combinations, one optimized and one not. Linear transmission spectra were calculated by 3D-FDTD calculations carried out with the Lumerical commercial software package[20] and the nonlinear response was calculated by the same package, based on the NL-3D-FDTD routines (For details of both the experiments and the calculations see the Methods section below).

**Linear transmission.** The linear transmission spectra of rectangular cavities, calculated and measured for a range of AR, are shown in Figure 1 a,b. As the aspect ratio increases from AR=1 to AR=4, two separate peaks evolve, representing the two accessible modes. The dominant mode, which strongly depends on the AR with transmission peak that changes continuously from 650 to 1100 nm, and a weaker one which hardly depends on the aspect ratio and remains around 650 nm for the entire range of aspect ratios. These two resonances originate with different physical mechanisms. The resonance at longer wavelengths is attributed to localized surface plasmons (LSP) excitation on the nanoholes and therefore strongly depends on their AR [21, 22], and like for metallic nanorods, the LSP resonance is red-shifted for more elongated rectangles. The observed peak around 650 nm is commonly referred to as the Fabry-Perot resonance [23, 24]. Figure 1c depicts three cross-sections at AR=1.0; 2.1; 3.6 clearly showing the optimality of AR=2.1 for resonance excitation at 800 nm. Note that for AR=1.0 the two modes seem to coalesce, and only a single peak with a slight shoulder is visible.

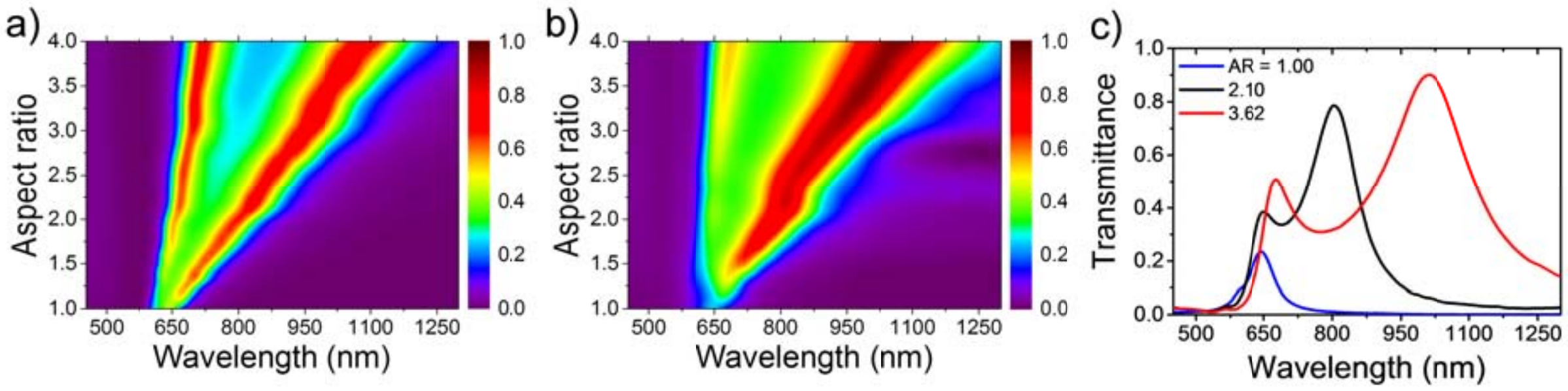

**FIGURE 1 -** Transmission spectra for different aspect ratios (AR): (a) calculated and (b) measured spectra covering a wide range of AR; (c) Three specific measured spectra for AR=1.0 , 2.1 and 3.6 . The calculations were done with coherent white light excitation (see text)

Several comments are in order:

1. Nanocavities differ from metallic rods in that due to Babinet's principle, equivalent behavior is seen for polarizations that are rotated by 90 degrees[25]. Thus, in both the calculation and measurements, the light was polarized along the short axis of the rectangle.
2. The transmission spectra were measured with white light which is spatially and temporally incoherent, whereas, in anticipation of the FWM experiment, the calculations presented here were performed with spectrally and spatially coherent light.
3. All the calculations presented in figure 1 were carried out for a single "perfect" cavity with perfect walls and theoretically sharp corners. Furthermore, the method of calculation is based on a single cavity with periodic boundary conditions, which thus dictates perfectly reproducible fabrication without any variability between the holes.

A more realistic calculation should allow for cavity dimension variability and less than perfect rectangle corners, and for incoherent light excitation. The numerical approach for performing such a calculation and some typical results are discussed in Supplementary Figure 1.

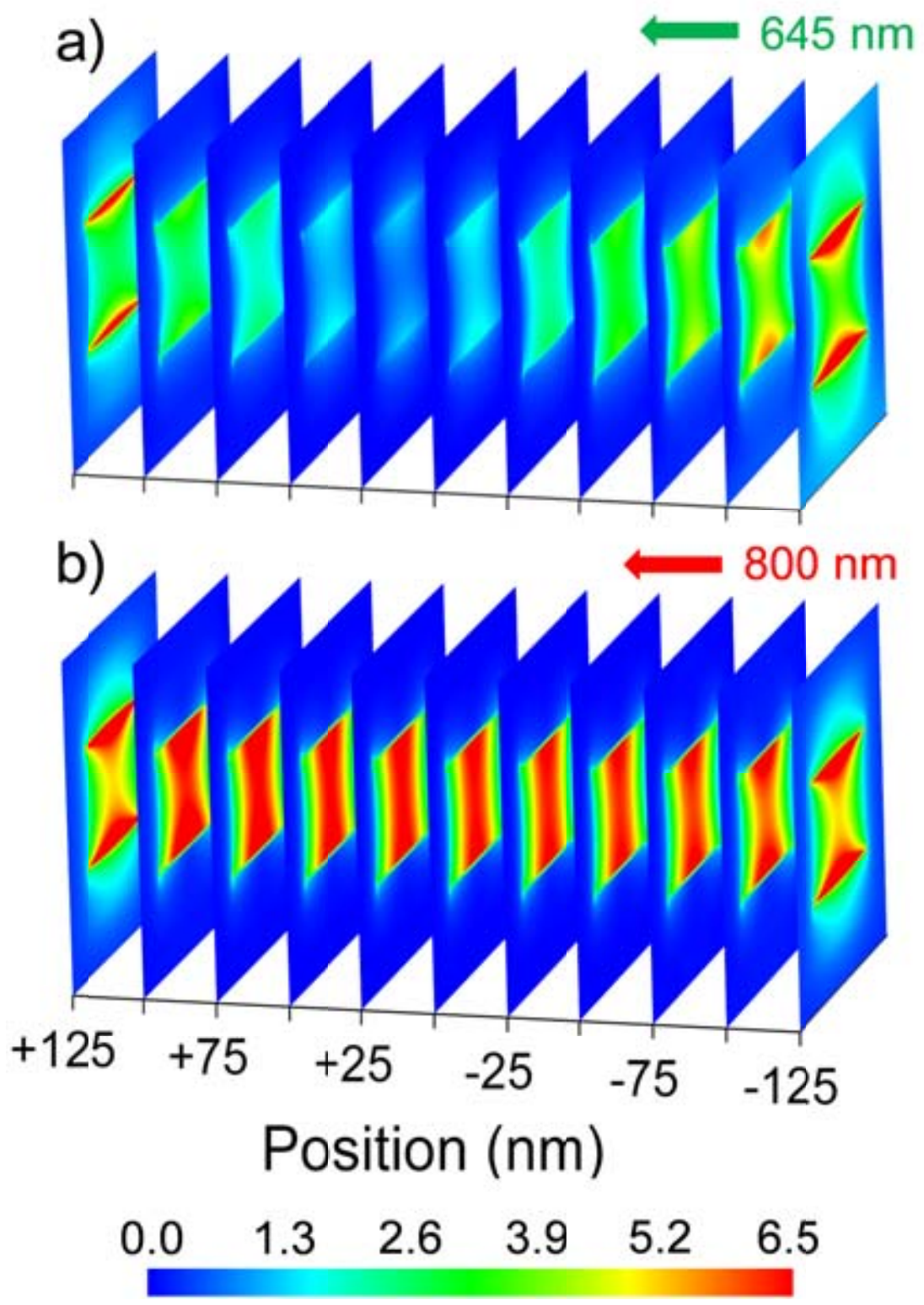

**FIGURE 2 -** Field distribution for the 645 nm and 800 nm modes propagating down a rectangular nanocavity with dimensions 290x138 $nm^2$ (AR=2.1) in a gold film of thickness of 250 nm.

To better visualize the two separate resonances, the calculated distributions of the electric fields at 650 nm and at 800 nm inside the nanocavity are depicted in figure 2 for a cavity of 290x138 $nm^2$ (AR=2.1). In both cases, light propagates from right to left, and the field distribution profiles (at 645 and at 800nm) are displayed for increasing propagation distance into the cavity. The difference between the two modes is clear. The FP mode around 645 nm (Fig. 2a) contains a node in the center of the cavity and field maxima at the ends, which is typical to an open-ended

Fabry-Perot resonator[23, 24]. The LSP mode at 800 nm propagates along the cavity with small variations in its transverse profile and shows strong frequency dependence (see figure 2b) on the AR. The two propagating modes are visualized in a 3D movie in Supplementary Movie 1

**Nonlinear Four Wave Mixing.** Using the results of the linear transmittance as the starting point, we formulate a strategy to obtain efficient FWM signal in the visible. The FWM signal (see illustration in figure 3) results from the coherent interaction between the two beams via the third-order nonlinear susceptibility $\chi^{(3)}$:

$$P^{(3)}(\omega_{FWM}, \omega_1, \omega_2) = \varepsilon_0 \chi^{(3)}(\omega_{FWM}, \omega_1, \omega_2)[E_1(\omega_1)]^2 E_2^*(\omega_2) \qquad (1)$$

Where $P^{(3)}$ is the third-order nonlinear polarization induced in the material and the intensity of the signal is given by $I_{FWM} \propto \left|P^{(3)}\right|^2 = {\varepsilon_0}^2 \left|\chi^{(3)}\right|^2 |E_1|^4 |E_2|^2$. Thus, the product $|E_1|^4 |E_2|^2$ of the input electric fields must be maximized. If we limit ourselves to only a single parameter of optimization, namely the AR, the LSP resonance can be tuned to match one input frequency, and since the dominant contribution comes from the quadratic dependence at $\omega_1$, optimal tuning means matching the LSP resonance to this frequency. We aim for a strong FWM signal in the visible, and this can be achieved at 645 nm, matching the Fabry-Perot mode to obtain enhanced EOT. As shown in the linear calculations and measurements, the LSP resonance should be tuned to match our strong beam at 800 nm (the fundamental frequency of the Ti:Sapphire laser), and this happens for AR = 2.1 . The fundamental input frequency $\omega_1 =$ 800 nm implies that the second input frequency should be $\omega_2 = 2\omega_1 - \omega_{FWM}$ =1065 nm.

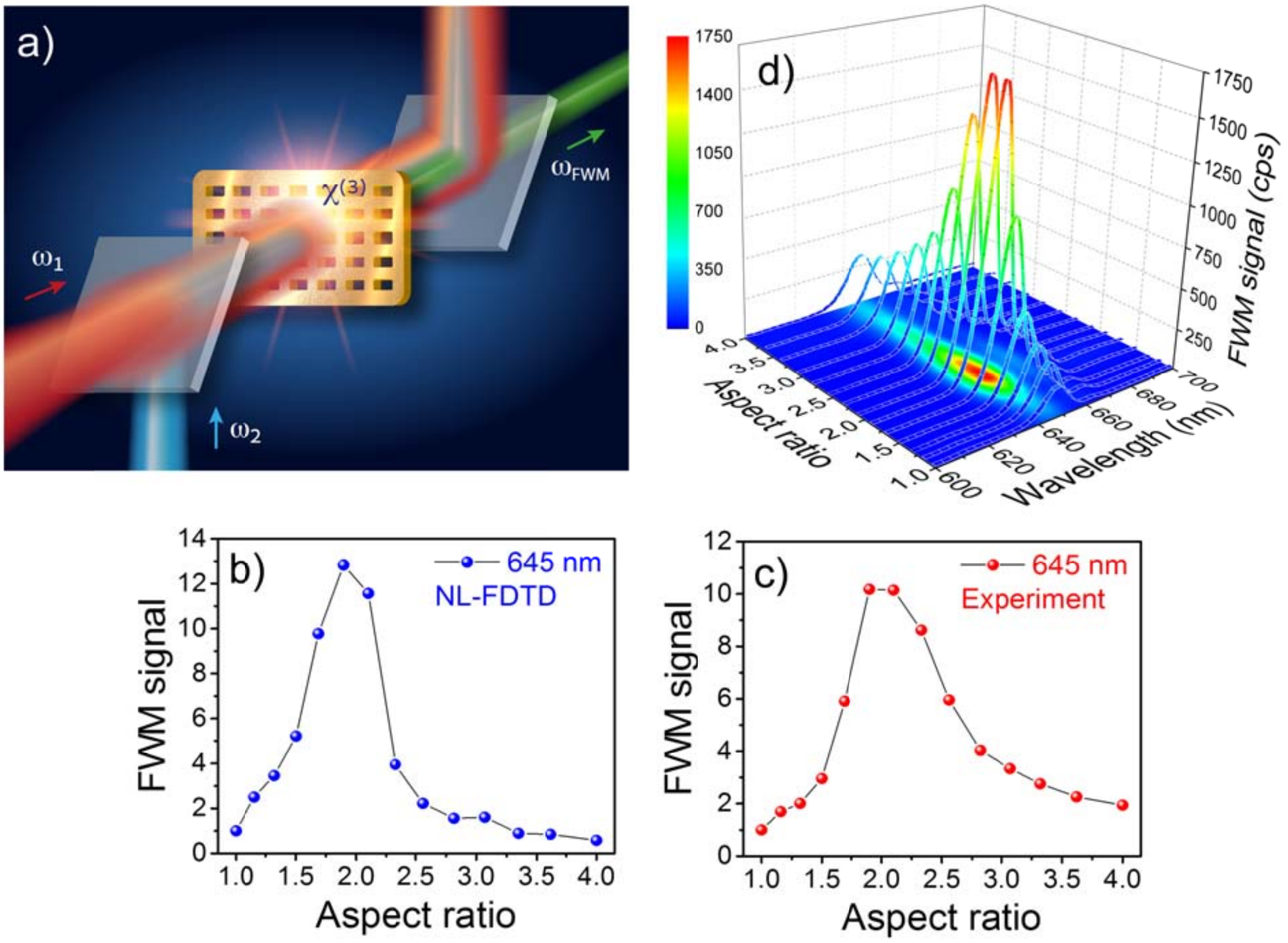

**FIGURE 3** - (a) Illustration of the FWM processes in an array of rectangular nanoholes. (b) Calculated and (c) measured AR dependence of the FWM signal at 645 nm and the corresponding (d) experimental spectra of the signal.

With the input frequencies determined, we calculate the expected FWM signal for different AR. We use an implementation of a NL-3D-FDTD method using the commercial Lumerical Solutions software[20]. The field update equations of the FDTD algorithm are modified to include a third-order NL polarization induced in a material with instantaneous $\chi^{(3)}$. Two temporally overlapped 60 fs plane waves centered at frequencies $\omega_1$ and $\omega_2$ were used as the FDTD sources to illuminate a single nanorectangle drilled in a gold film with periodic boundary conditions. The $\chi^{(3)}$ of gold was chosen according to the value reported in literature[26] in this spectral region ($\chi^{(3)}$ =1x10$^{-18}$ m$^2$/V$^2$). The electric field amplitude of both sources was set to 1x10$^9$ V/m (equivalent to peak power of 6x10$^{11}$ W/cm$^2$ or pulse energy of 80 nJ for our pulses and geometry) and the fields are polarized along the short axis of the rectangle. The *z* component of the Poynting vector is spatially integrated on two z-normal planes positioned after (forward FWM) and before (backward FWM) the Au film. The results of this model calculation are depicted in figure 3 where the spectrally integrated FWM signal at $\omega_{FWM}$ = 645 nm is plotted against the aspect ratio for input frequencies $\omega_1$ =800 nm; $\omega_2$ = 1065 nm. A clear resonance is predicted near AR=2.1 with a very rapid drop for other values, smaller or larger. In particular, a rather weak signal is predicted for a square cavity of AR=1.

Next, we verified these predictions experimentally. We used two 60 fsec pulses, one centered at the fundamental Ti:Sapphire frequency $\omega_1$ = 800 nm, and $\omega_2$ that was obtained from a tunable optical parametric amplifier (OPA) with frequency ranging between 1000 nm to 1300 nm with the same pulse duration and repetition rate (1 KHz). The $\omega_1$ and $\omega_2$ beams are spatially and temporally overlapped at the free-standing gold film in which the nanohole arrays are milled (see Figure 3). The FWM signal is filtered and measured in both forward and backward directions with a CCD detector coupled to a spectrometer. The use of a free standing film, in spite of the difficulty in preparing and handling it, is critical for the removal of the much stronger coherent, nonresonant FWM emission from a thick glass substrate. Furthermore, in a free standing film, the matching of the refractive index at both interfaces of the arrays[27] has the added advantage of increasing the observed optical transmittance, higher than >80% in the arrays with large AR. Having similar interfaces on both sides of the film also gives rise to enhanced fields near the surface, but this is outside the scope of the present work.

In the first set of FWM experiments, the wavelength of the OPA was tuned to $\omega_2$=1050 nm so that the generated FWM signal at $\omega_{FWM} = 645$ nm will match the FP cavity resonance which was predicted theoretically and observed experimentally in the linear transmission measurement. Figure 3 shows the spectrally integrated FWM signal for several different AR along with the corresponding spectra of the FWM signal around 645 nm. As predicted by the NL calculations, shape dependent resonance is seen for rectangles with AR = 2.1, with more than an order of magnitude enhancement for AR=2.1 as compared to AR=1.

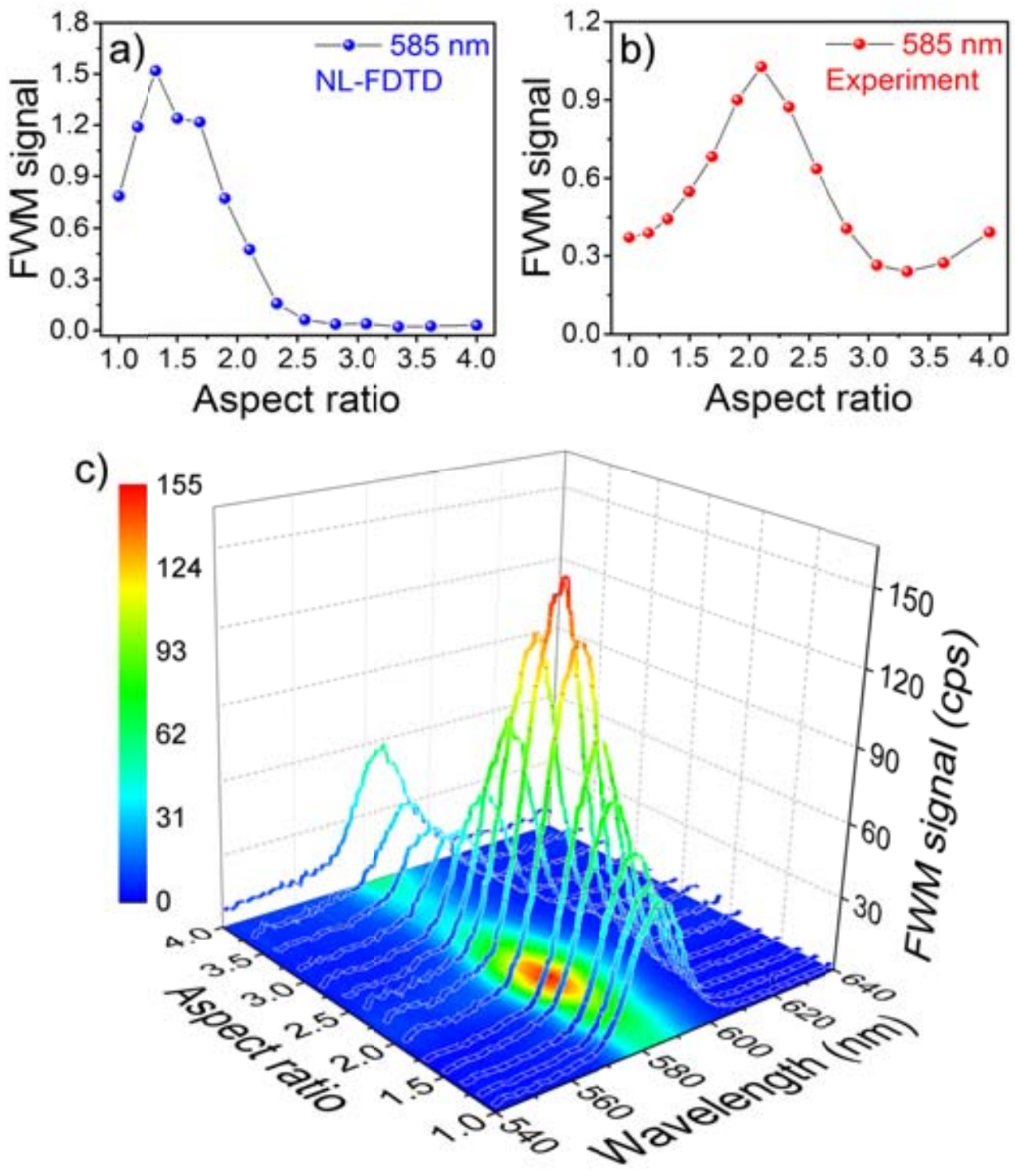

**FIGURE 4 -** (a) Calculated and (b) measured AR dependence of the FWM signal at 585 nm. (c) Measured spectra of the FWM signal around 585 nm.

To clarify the significance of the FP-resonance for the NL generation, a different input wavelength was chosen ($\omega_2 = 1265$ nm) to generate an unoptimized signal at $\omega_{FWM} = 585$ nm, which is spectrally removed from the FP resonance. Figure 4 depicts the calculated and measured AR dependence of the integrated signal. While the FWM signal is still maximal at AR = 2.1, the intensity of the FWM signal at 585nm is a full order of magnitude weaker. As discussed, the maximum at AR=2.1 can be traced directly to the LSP resonance at 800nm, and therefore it is seen in both the optimized FWM at 645nm, and at the non-optimized FWM at 585nm. As seen in the linear transmission spectra, and supported by the 3D-FDTD calculations, the field enhancement at $\omega_2 = 1265$ nm is similar to that for $\omega_2 = 1050$ nm (both frequencies are far out of resonance with the LSP mode for AR=2.1). Therefore, the contribution of the EMF enhancement in eq. (1) is similar for both $\omega_2$ frequencies. Assuming a smooth dispersion relation for $\chi^{(3)}$, the enhanced FWM at 645 nm may be directly attributed to EOT at the FP resonance.

We have also measured the backwards FWM, and found it to be more than an order of magnitude smaller, because under our experimental conditions, phase matching is more efficient in the forward direction[18]. In all these measurements, the signal from the bare gold film (without holes) is either not visible at all or hardly detectable above the noise level. Note also that the measured shape resonance is broader than the calculated one, its contrast ratio less pronounced and the peak is displaced to slightly lower aspect ratio. These observations are discussed below.

As a last point in this section, we try to estimate the FWM efficiency. When detector efficiency and losses in optical components along the detection path of the signal (estimated to be around 90%) are taken into account, the extracted FWM conversion efficiency is of the order of $10^{-8}$, which is comparable to the published efficiency of a BBO nonlinear crystal when calculated for the same thickness of 250 nm[28]. This result is promising as the current first attempt to design efficient nonlinear response was based on optimizing only a single parameter, namely the AR. A large number of other parameters can and will be considered, including size and shape of the cavities, fine fabrication details, spacing between them, film properties such a thickness, etc. These optimizations will be the subject of a future publication.

## **DISCUSSION**

The current work is a demonstration that rational design of metallic nanocavities can lead to enhanced, efficient Four Wave Mixing. When designing a strategy for optimizing the shape of nanocavities in a thin film, one may consider many different parameters as variables. These include, among others, film thickness and composition, cavity size and shape (rectangles, triangles, circles or other simple geometrical shapes), shapes with more complex symmetries, coupled cavities of different degree of coupling, and quite a few other parameters. Like many other examples, optimization in a multi-dimensional space is rather complex and one often resorts to computer algorithms and loses physical insight. For this reason we decided to limit ourselves to a single, physical parameter so that the optimization process can be followed and analyzed. Thus, we have focused on rectangles, and optimized their aspect ratio. However, for completeness, we illustrate the very strong dependence of the FWM efficiency on parameters other than the aspect ratio, which in turn can be used for optimization at other wavelengths or under different experimental conditions. Figure 5 depicts a calculated optimization of the FWM amplitude at 585nm for arrays of rectangles with varying periodicity and aspect ratio. Clearly, very pronounced dependencies are observed, with the optimal spacing being around 590 nm, and the optimal AR being 2.1 respectively. The FWM amplitude attained with these parameters is approximately 6-fold larger than what was achieved with the previous set of parameters used for the optimization at 645 nm. Note, however, that real optimization must be done in a multidimensional space, optimizing all the parameters simultaneously, and not one parameter at a time, which may only lead to local maxima.

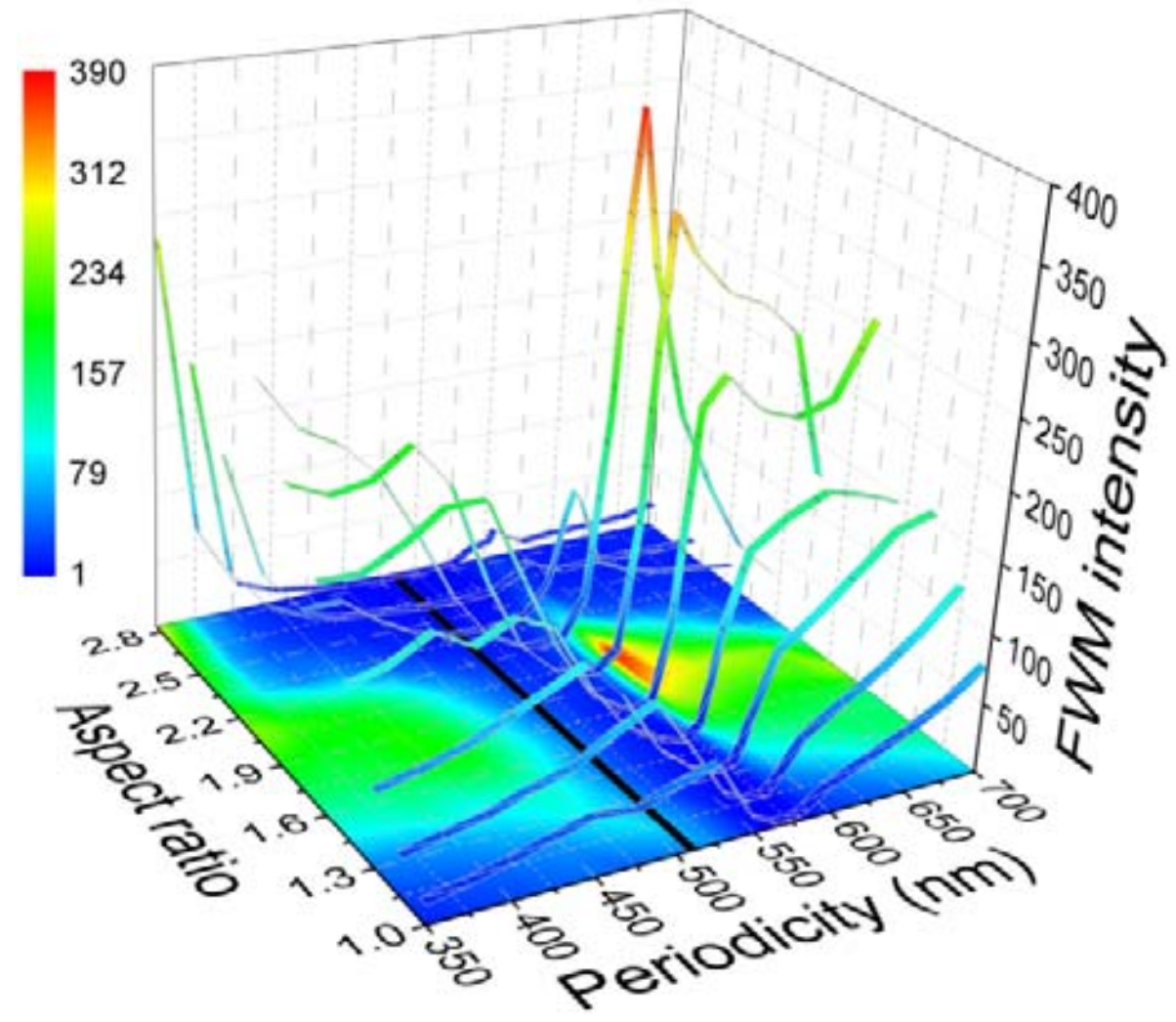

**FIGURE 5** - Dependence of the FWM signal at 585 nm on the aspect ratio and the periodicity. For periodicities between 520 and 550 nm, Wood's anomaly tends to decrease the FWM signal. The black line indicated the periodicity of 510 nm used in our experiments.

In a related context, we have previously shown that triangular nanocavities of optimized size offer strong enhancement of SHG[13], and have investigated the coupling of such triangles over large distances[29]. To the best of our knowledge, similar studies have not been reported for higher order nonlinearities [5, 10].

In the NL calculations, the FWM resonance appears shifted to slightly smaller AR, and the peak-to-baseline ratio is larger than the experimental value. These differences may be caused by inaccuracies in the determination of fabrication parameters such as film thickness and nanocavities' exact dimensions and shape, FIB resolution, and in particular sharpness (or roundedness) of edges and corners that tends to decrease the field enhancement. To illustrate the influence of these effects and sensitivity of the resonance to exact parameters, in Figure 6 we show the calculated FWM signal AR dependence for different size cavities. Not only does the peak-to-baseline ratio change, but also the FWM resonance is shifted, depending on the cavity size. Similar observations were previously reported for SHG[10].

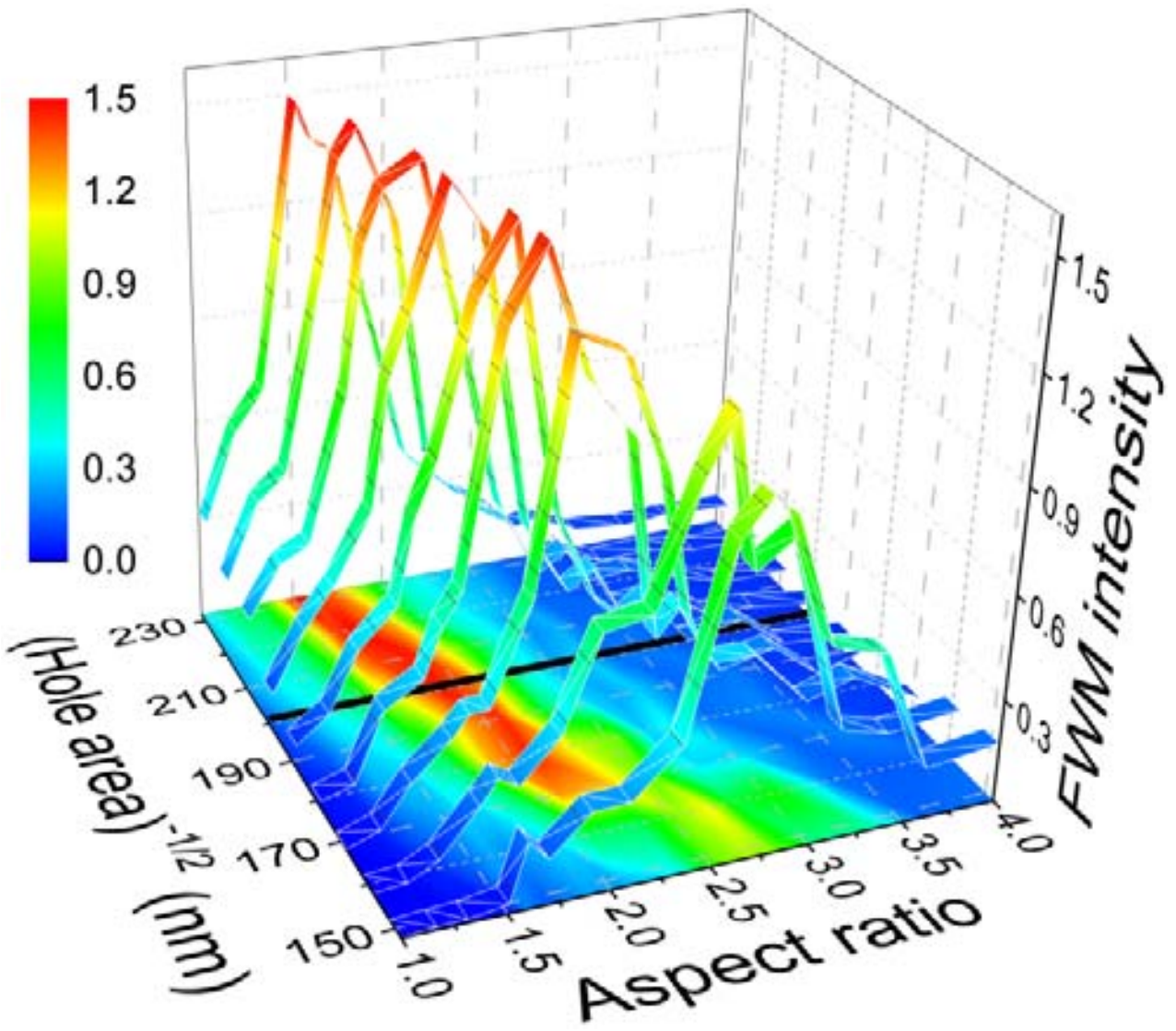

**FIGURE 6 -** Calculated AR dependence of the integrated FWM signal at 645 nm for different areas defined by a square hole with the shown size. The black line is for 200x200 nm$^2$ used in our experiments

The FWM shape resonance may be investigated analytically using a nonlinear coupled mode theory for energy exchange among the longitudinal modes inside the cavities (Supplementary discussion 1). The analysis is an extension to FWM of the formalism presented in Ref. [10] for SHG, and as in reference 10, the generated FWM intensity depends on many factors such as: mode-overlap, phase matching, field enhancement at the input frequencies and attenuation upon propagation of all waves involved. In spite of some inherent limitations, the model provides a semi quantitative analysis and captures the underlying physical mechanism of FWM generation inside nanocavities. As the incoming waves propagate through the cavity, it is found, by the numerical solutions of the propagating waves, that the fields are much stronger inside them than on the surface so that most of the nonlinear interaction takes place, and most of the FWM signal is generated inside the nanoholes. Two longitudinal modes propagate inside the nanoholes, which behave like waveguides. As they propagate, the modes interact with the metal and exchange energy with the FWM mode which builds up. The total generated forward FWM signal is assumed to be sum of the intensity of the FWM modes at the cavity exit.

As anticipated, the predominant factor for the FWM shape resonance is the field enhancement at $\omega_1 = 800$ nm, which exhibits a resonance for AR = 2.1. Although there is a shape resonance in the field enhancement at $\omega_2 = 1050$ nm also for AR = 3.6, it contributes to the FWM signal with the square of the field amplitude, as compared to the stronger, fourth power amplitude dependence of the EF enhancement at $\omega_1 = 800$ nm.

In summary, we demonstrated efficient FWM generation in arrays of periodic nanoholes drilled in a gold film. The FWM efficiency, which is comparable to the conversion efficiency of good nonlinear optical crystals, can be optimized by rational design of the geometry. To demonstrate the utility of the approach, we opted to control only a single parameter, the aspect ratio of the rectangles, and showed that when the cavity resonances match the frequencies involved in the FWM process, significant enhancement is observed. NL-3D-FDTD calculations were implemented that could predict, with good agreement, the shape FWM resonance observed latter in the experiments. A nonlinear coupled mode model was used to explain the FWM generation and the shape resonance in terms of the various individual physical processes that contribute to the FWM generation. We have discussed other parameters that can be optimized, and have included one such example. The control of the processes that lead to strong NL generation in very thin metamaterials is fundamental for the development of the next generation of optical devices, and more work is required in order to explore the very large parameter space available for optimization of specific processes.

## METHODS

### Sample Preparation

Large area, high quality free standing gold films were prepared by depositing gold on silicon, and etching the silicon from under the gold (Supplementary Figure 1). The films were characterized by SEM and found to be of very high surface quality. The rectangular nanocavities were milled by a focused ion beam machine (*FEI, Helios Nano Lab 600i*). Various aspect ratio rectangles were milled (Figure 7), with a constant spacing of 510 nm between them, and for all the arrays, the milled area was kept constant and equal to that of a 200x200 nm$^2$ square.

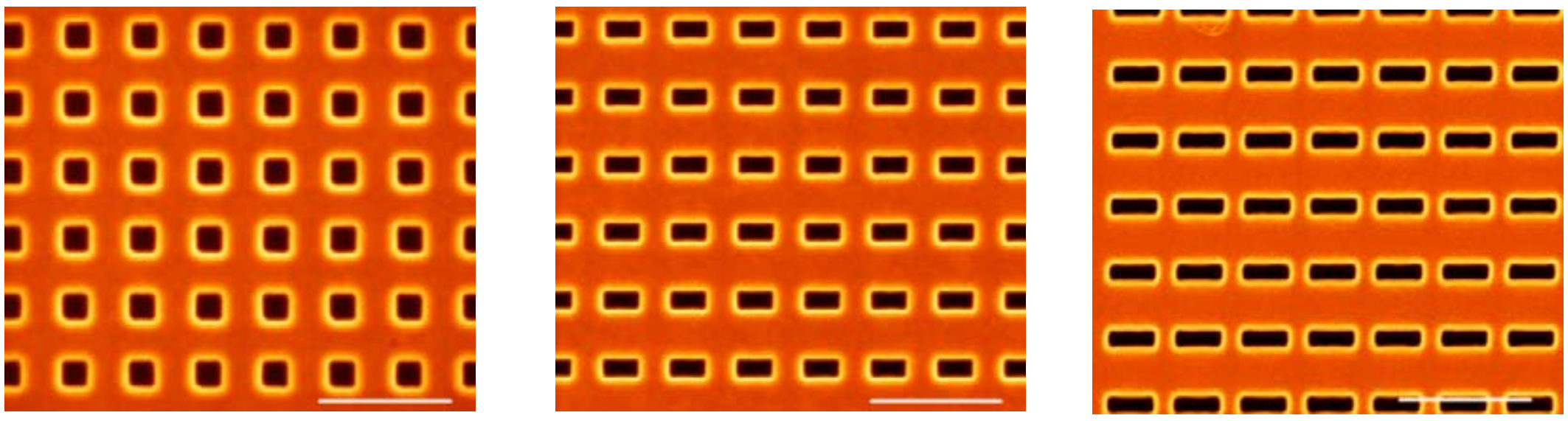

**FIGURE 7 -** SEM pictures of arrays of rectangles milled in a free standing, 250 nm thick gold film. Three different aspect ratios are shown (from left to right) AR=1, AR=1.5 and AR=3.1, the scale bar is 1 µm.

## Linear Transmission spectral measurements

The linear transmittance measurements were made using a 20 W Quartz-Tungsten-Halogen Lamp (Newport, 66310 QTH) as the light source. The measurements were carried out for normally incident light. Due to the limited spectral responsivity of the photodector, the spectra were taken in two different spectral ranges. From 450 to 900 nm, the light transmitted through the arrays was detected by a CCD (Jobin-Yvon, Symphony) coupled to a $f$ = 190 mm spectrometer (Jobin-Yvon, Triax 180). From 900 to 1350 nm, the transmitted light was modulated and lock-in detected by an InGaAs photodiode (Thorlabs, SM05PD5A) coupled to a spectrometer (SPEX) controlled by a computer. The spectra acquired with the different detection systems were each normalized to the light transmitted through an open area equal to the integrated area of the exposed array, and stitched for complete spectral visualization.

## Linear 3D-FDTD calculations

The 3D-FDTD method (Lumerical Solutions [20] software package) was used to calculate the linear transmission spectra of the arrays. The dimensions of the fine mesh around the rectangles were set to dx = 2 nm, dy = 2 nm and dz = 5 nm. Here, x, y are the long and short axes of the rectangles, and z is the propagation direction. Periodic boundary conditions were used along the x and y axis while perfectly matched layers were used in the z axis. The light source (plane wave) was polarized along the y axis and traveled with wavevector parallel to the z axis. The dielectric constants of the gold metal were taken from the Palik table[30]. Anticipating the coherent FWM experiments, in the calculation we assume a coherent source also for the linear transmission calculation. However, since most transmission spectral measurements (including ours) are performed with an incoherent white light source, there are differences between the observed and calculated spectra, and these are discussed in Supplementary Figure 1

**Four Wave Mixing**

In the FWM experiment two pulses are used as inputs: the fundamental at $\omega_1 =$ 800 nm, generated by a Chirp Pulse Amplified Ti:Sapphire laser (Spectra Physics Spitfire pumped by a Mai Tai), and a pulse from an optical parametric amplifier (OPA) pumped by the same laser and tunable between 1000 nm to 1300 nm. Both pulses operate at a repetition rate of 1 KHz, they are 60 fsec long, of power 68 and 12 μJ respectively. The input pulses are spatially and temporally overlapped at the nanohole arrays. The FWM signal is generated at $\omega_{FWM} = 2\omega_1 - \omega_2$, the signal is filtered and measured in both forward and backward directions with a CCD detector coupled to a spectrometer (See supplementary figure 3).

**Nonlinear 3D-NL-FDTD calculations**

The 3D nonlinear finite differences time domain (3D-NL-FDTD) simulations were carried out with the same Lumerical Solutions software. The FDTD method solves Maxwell's equations in discrete time and space (Yee's cell). As an illustration of the method, consider the simplest 1D case of a TM field propagating along the z direction in a nonmagnetic, nondispersive and homogenous medium. The Maxwell equations are written as:

$$\mu_0 \frac{\partial H_x}{\partial t} = \frac{\partial E_y}{\partial z} \tag{1}$$

$$\frac{\partial D_z}{\partial t} = \frac{\partial H_x}{\partial z} \quad (2)$$

where $E$ and $H$ are the electric and magnetic fields respectively, $\mu_0$ is the magnetic permeability and $D_y = \varepsilon E_y$ is the displacement field with the frequency-independent electric permittivity $\varepsilon$. The previous equations can be discretized using the Yee's algorithm:

$$H_x^{n+1/2}\left[m+\frac{1}{2}\right] = H_x^{n-1/2}\left[m+\frac{1}{2}\right] + \frac{\Delta t}{\mu_0 \Delta z}\left(E_y^n[m+1] - E_y^n[m]\right) \quad (3)$$

$$D_y^{n+1}[m] = D_y^n[m] + \frac{\Delta t}{\Delta z}\left(H_x^{n+1/2}\left[m+\frac{1}{2}\right] - H_x^{n+1/2}\left[m-\frac{1}{2}\right]\right) \quad (4)$$

$$\varepsilon E_y^{n+1}[m] = D_y^{n+1}[m] \quad (5)$$

where $n$ is the time step and $m$ denotes the spatial position of a single cell. The third-order nonlinearity of the material is added in the dielectric constant $\varepsilon = \varepsilon_0 + \chi^{(3)}E^2$, where $\chi^{(3)}$ is the (nondispersive) third-order susceptibility. Therefore, by modifying the dielectric constant of the material to include the nonlinearity, we can obtain a new update equation for the electric field:

$$E_y^{n+1}[m] = \frac{D_y^{n+1}[m]}{\varepsilon_0 + \chi^{(3)}\left(E_y^n[m]\right)^2} \quad (6)$$

The electric field calculated in the previous time step is used to update the new electric field, which in turn is used to update the new magnetic field (eq. 4) therefore closing the loop. To write eq. 6, the nonlinearity is assumed instantaneous.

In the 3D-NL-FDTD simulations, the dimensions of the mesh around the rectangles were set to dx = 6 nm, dy = 4 nm and dz = 5 nm. We simulate a single rectangle milled in a 250 nm thick gold film and chose periodic boundary conditions with period 510 nm in both x and y directions. Perfectly matched layers were added in the z direction. The dielectric constants were taken from Palik. The diagonal terms of the nonlinear susceptibility were set to $\chi^{(3)} = 1\text{x}10^{-18}$ $m^2/V^2$ and the nondiagonal to zero. A y-polarized plane wave centered at $\omega_1 = 800$ nm with pulse duration 60 fs, offset 120 fs and amplitude $1\text{x}10^9$ V/m is mixed with another plane wave centered at $\omega_2 = 1050$ nm of same pulse duration, offset and amplitude. The two plane waves travel along the z direction and are initially set 150 nm far from the surface. The forward (backward) signal centered at $\omega_{FWM} = 645$ nm is acquired by spatially integrating the z-component of the Poynting vector in a power monitor located after (before) the metal surface.

## ACKNOWLEDGEMENT

Many useful discussions with Adi Salomon in early stages of this work are thankfully acknowledged. We are grateful to Alex Yoffe for critical contributions and help with fabrication of the free standing gold film. Discussions with Guy Shalem and help with the numerical simulations are thankfully acknowledged. This work was funded, in part, by the Israel Science Foundation, by the ICORE program, by an FTA grant on INNI, the Israel National Nano Initiative, and by a grant from the Leona M. and Harry B. Helmsley Charitable Trust

## REFERENCES

[1] N. F. Yu and F. Capasso, "Flat optics with designer metasurfaces," *Nature Materials,* vol. 13, pp. 139-150, Feb 2014.

[2] J. A. Schuller, E. S. Barnard, W. S. Cai, Y. C. Jun, J. S. White, and M. L. Brongersma, "Plasmonics for extreme light concentration and manipulation," *Nat Mater,* vol. 9, pp. 193-204, Mar 2010.

[3] L. Novotny and N. van Hulst, "Antennas for light," *Nature Photonics,* vol. 5, pp. 83-90, Feb 2011.

[4] M. Kauranen and A. V. Zayats, "Nonlinear plasmonics," *Nature Photonics,* vol. 6, pp. 737-748, Nov 2012.

[5] J. A. H. van Nieuwstadt, M. Sandtke, R. H. Harmsen, F. B. Segerink, J. C. Prangsma, S. Enoch*, et al.*, "Strong modification of the nonlinear optical response of metallic subwavelength hole arrays," *Physical Review Letters,* vol. 97, Oct 6 2006.

[6] P. Genevet, J. P. Tetienne, E. Gatzogiannis, R. Blanchard, M. A. Kats, M. O. Scully*, et al.*, "Large Enhancement of Nonlinear Optical Phenomena by Plasmonic Nanocavity Gratings," *Nano Letters,* vol. 10, pp. 4880-4883, Dec 2010.

[7] S. Kim, J. H. Jin, Y. J. Kim, I. Y. Park, Y. Kim, and S. W. Kim, "High-harmonic generation by resonant plasmon field enhancement," *Nature,* vol. 453, pp. 757-760, Jun 5 2008.

[8] J. Lee, M. Tymchenko, C. Argyropoulos, P. Y. Chen, F. Lu, F. Demmerle*, et al.*, "Giant nonlinear response from plasmonic metasurfaces coupled to intersubband transitions," *Nature,* vol. 511, pp. 65-U389, Jul 3 2014.

[9] B. Lamprecht, A. Leitner, and F. R. Aussenegg, "Femtosecond decay-time measurement of electron-plasma oscillation in nanolithographically designed silver particles," *Applied Physics B-Lasers and Optics,* vol. 64, pp. 269-272, Feb 1997.

[10] B. L. Wang, R. Wang, R. J. Liu, X. H. Lu, J. M. Zhao, and Z. Y. Li, "Origin of Shape Resonance in Second-Harmonic Generation from Metallic Nanohole Arrays," *Scientific Reports,* vol. 3, Aug 5 2013.

[11] Y. Zhang, N. K. Grady, C. Ayala-Orozco, and N. J. Halas, "Three-Dimensional Nanostructures as Highly Efficient Generators of Second Harmonic Light," *Nano Letters,* vol. 11, pp. 5519-5523, Dec 2011.

[12] P. Schon, N. Bonod, E. Devaux, J. Wenger, H. Rigneault, T. W. Ebbesen*, et al.*, "Enhanced second-harmonic generation from individual metallic nanoapertures," *Opt Lett,* vol. 35, pp. 4063-4065, Dec 1 2010.

[13] A. Salomon, A. Zielinski, R. Kolkowski, J. Zyss, and Y. Prior, "Shape and Size Resonances in Second Harmonic Generation from Plasmonic Nano-Cavities " *J. Phys. Chem. C,* vol. 117, pp. 22377-22382, 2013.

[14] Y. Zhang, Y. R. Zhen, O. Neumann, J. K. Day, P. Nordlander, and N. J. Halas, "Coherent anti-Stokes Raman scattering with single-molecule sensitivity using a plasmonic Fano resonance," *Nature Communications,* vol. 5, Jul 2014.

[15] J. Renger, R. Quidant, N. van Hulst, and L. Novotny, "Surface-Enhanced Nonlinear Four-Wave Mixing," *Physical Review Letters,* vol. 104, Jan 29 2010.

[16] Y. Zhang, F. Wen, Y. R. Zhen, P. Nordlander, and N. J. Halas, "Coherent Fano resonances in a plasmonic nanocluster enhance optical four-wave mixing," *Proceedings of the National Academy of Sciences of the United States of America,* vol. 110, pp. 9215-9219, Jun 4 2013.

[17] H. Harutyunyan, G. Volpe, R. Quidant, and L. Novotny, "Enhancing the Nonlinear Optical Response Using Multifrequency Gold-Nanowire Antennas," *Physical Review Letters,* vol. 108, May 23 2012.

[18] H. Suchowski, K. O'Brien, Z. J. Wong, A. Salandrino, X. B. Yin, and X. Zhang, "Phase Mismatch-Free Nonlinear Propagation in Optical Zero-Index Materials," *Science,* vol. 342, pp. 1223-1226, Dec 6 2013.

[19] T. W. Ebbesen, H. J. Lezec, H. F. Ghaemi, T. Thio, and P. A. Wolff, "Extraordinary optical transmission through sub-wavelength hole arrays," *Nature,* vol. 391, pp. 667-669, Feb 12 1998.

[20] Lumerical, "Lumerical solutions, Inc. http://www.lumerical.com/tcad-products/fdtd."

[21] K. J. K. Koerkamp, S. Enoch, F. B. Segerink, N. F. van Hulst, and L. Kuipers, "Strong influence of hole shape on extraordinary transmission through periodic arrays of subwavelength holes," *Phys Rev Lett,* vol. 92, May 7 2004.

[22] K. L. van der Molen, K. J. Klein Koerkamp, S. Enoch, F. B. Segerink, N. F. van Hulst, and L. Kuipers, "Role of shape and localized resonances in extraordinary transmission through periodic arrays of subwavelength holes: Experiment and theory," *Phys Rev B,* vol. 72, p. 045421, Jul 2005.

[23] S. Astilean, P. Lalanne, and M. Palamaru, "Light transmission through metallic channels much smaller than the wavelength," *Optics Communications,* vol. 175, pp. 265-273, Mar 1 2000.

[24] H. C. Guo, T. P. Meyrath, T. Zentgraf, N. Liu, L. W. Fu, H. Schweizer*, et al.*, "Optical resonances of bowtie slot antennas and their geometry and material dependence," *Optics Express,* vol. 16, pp. 7756-7766, May 26 2008.

[25] T. Zentgraf, T. P. Meyrath, A. Seidel, S. Kaiser, H. Giessen, C. Rockstuhl*, et al.*, "Babinet's principle for optical frequency metamaterials and nanoantennas," *Physical Review B,* vol. 76, Jul 2007.

[26] R. W. Boyd, Z. M. Shi, and I. De Leon, "The third-order nonlinear optical susceptibility of gold," *Optics Communications,* vol. 326, pp. 74-79, Sep 1 2014.

[27] A. Krishnan, T. Thio, T. J. Kima, H. J. Lezec, T. W. Ebbesen, P. A. Wolff*, et al.*, "Evanescently coupled resonance in surface plasmon enhanced transmission," *Optics Communications,* vol. 200, pp. 1-7, Dec 15 2001.

[28] eskmaoptics, "http://eksmaoptics.com/out/media/Femtoline(2).pdf."

[29] A. Salomon, Y. Prior, M. Fedoruk, J. Feldmann, R. Kolkowski, and J. Zyss, "Plasmonic Coupling between Metallic Nanocavities," *Journal of Optics,* vol. 16, p. 114012, 2014.

[30] E. D. Palik, *Handbook of Optical Constants of Solids* Academic press, San Diego, California, 1998.

# Supplementary information for

## Rational design of metallic nanocavities for resonantly enhanced four-wave mixing

Euclides Almeida and Yehiam Prior

*Department of Chemical Physics, Weizmann Institute of Science, Rehovot 76100, Israel*

**<u>Supplementary Figure 1</u>:** Non-uniform sample with spatially incoherent excitation

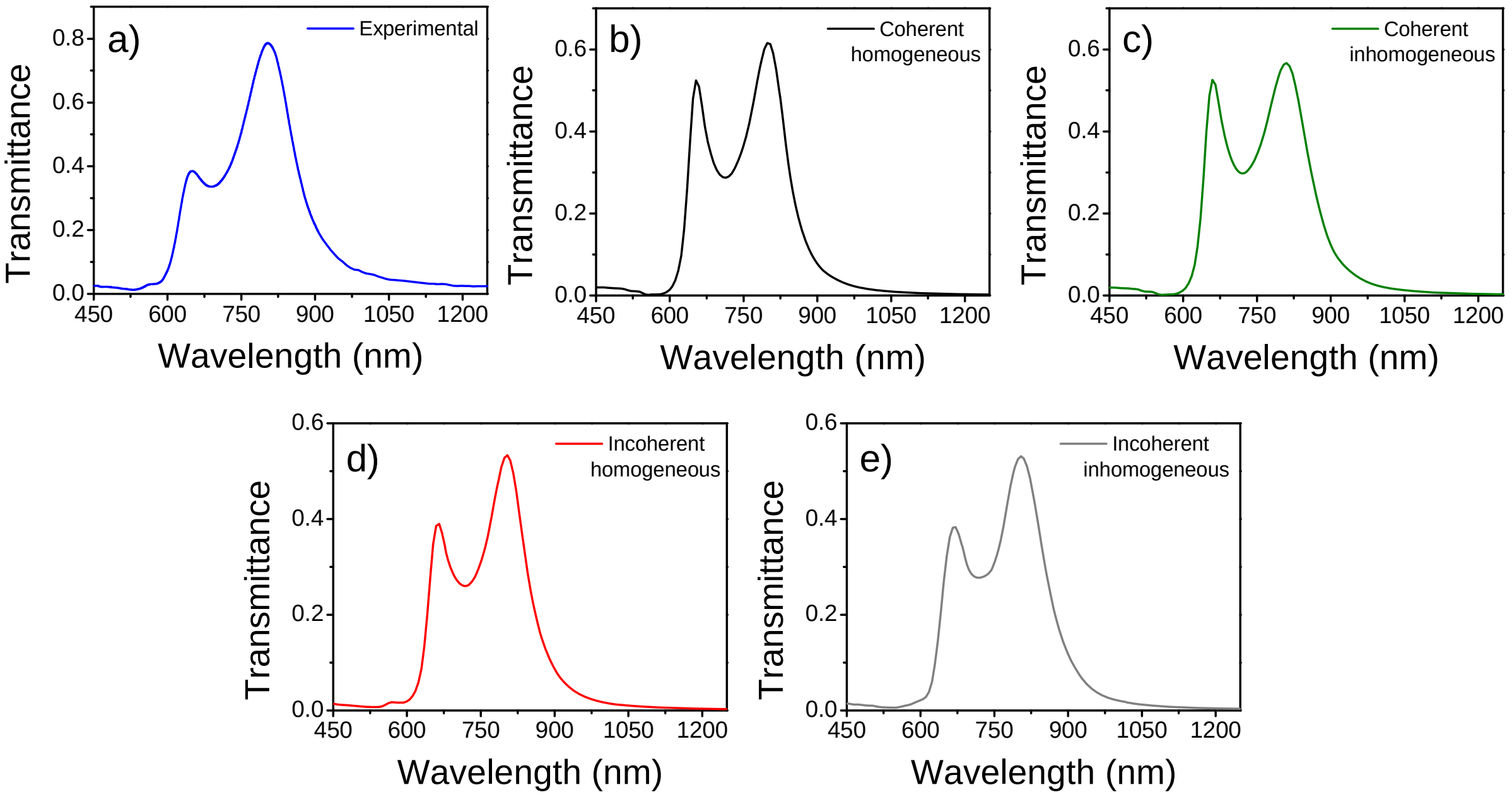

In anticipation of the coherent FWM experiments, and to simulate the laser sources used for these experiments, the linear transmission was calculated using spatially and temporally coherent light source. The linear measurements, however, are based on white light illumination with a source that is neither spatially nor temporally coherent. Furthermore, the calculations are performed on a single cavity, with periodic boundary conditions, which implies a perfect sample where all cavities are identical, obviously unachievable in experimental reality. To test the applicability of these approximations, we repeated some of the calculations for spatially incoherent light, and for nonuniform samples. The light source is simulated as follows: The light source is divided into 12x12 plane wave sources each one spans 2.1 μm in both x and y directions, each centered on a different hole and having a random phase. Sample nonuniformity is simulated by performing the calculation on an array with 12x12 holes. The periodicity of the square array is maintained at 510 nm, but the dimensions of each hole are allowed to randomly vary (with Gaussian distribution) within 7 nm in x and y around the nominal size of 290x138 $nm^2$ (AR = 2.1).

a) The experimental transmittance of an array of rectangles

b) Simulated transmittance for coherent excitation and homogeneous sample

c) Simulated transmittance for coherent excitation and inhomogeneous sample

d) Simulated transmittance for incoherent excitation and homogeneous sample

e) Simulated transmittance for incoherent excitation and inhomogeneous sample

Note that the Wood's anomaly, which is a result of long-range coherent interaction (interference), disappears for the simulated spectra of coherent excitation. Note also that for the inhomogeneous samples, linewidths of Fabry-Perot and plasmonic resonances are inhomogenously broadened.

**Supplementary Figure 2:** Free standing gold films

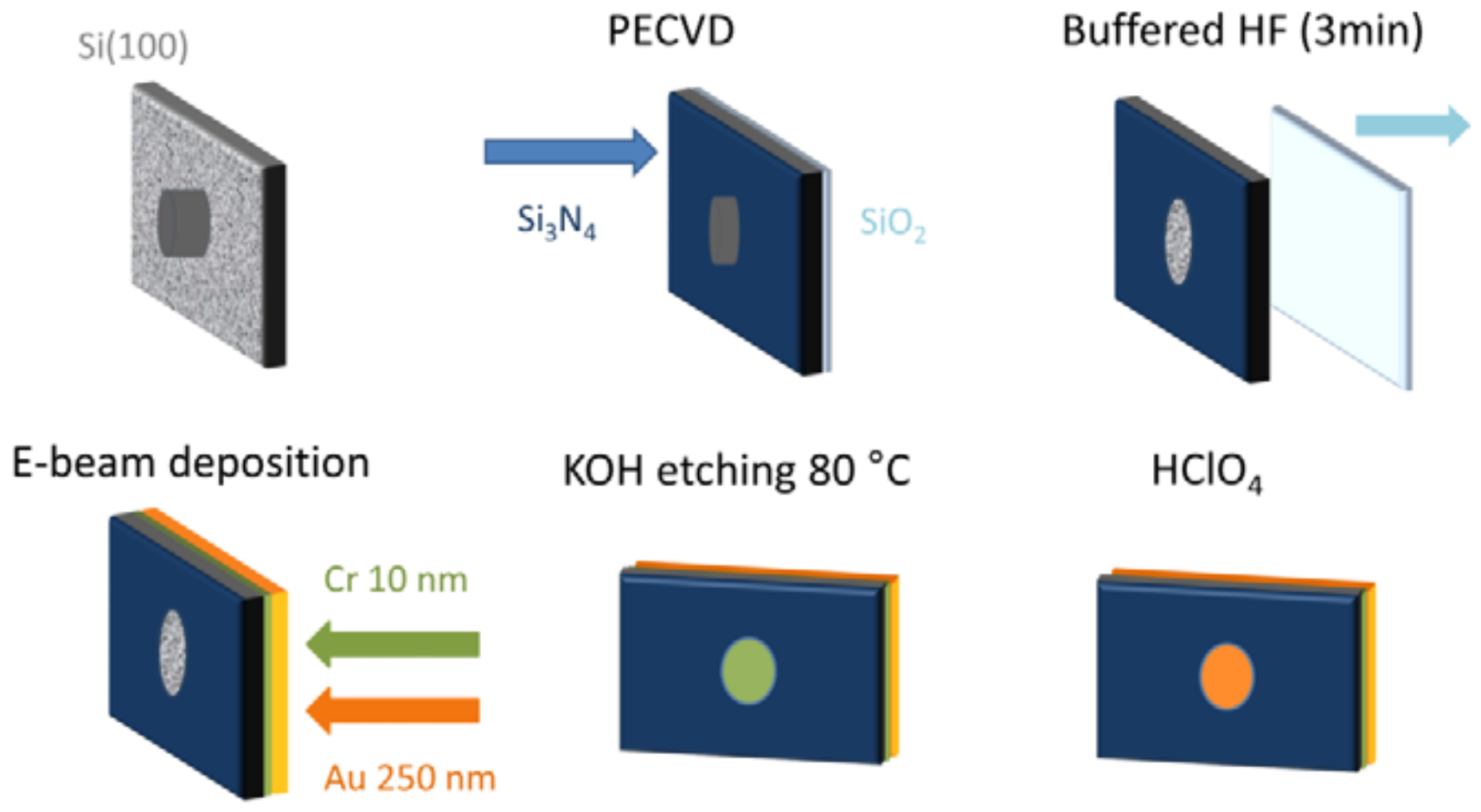

The preparation of free standing gold films was done out as follows: A 5 µm thick $Si_3N_4$ mask was deposited onto the rough side of a [100] silicon wafer by plasma-enhanced chemical vapor deposition (PECVD). The substrate was immersed in a buffered oxide etch solution for 3 minutes to remove the silica layer formed during the PEVCD process. A 10 nm thick chromium adhesion layer was deposited onto the smooth surface of the wafer by e-beam evaporation, followed by evaporation of a 250 nm thick gold layer. The substrate was then chemically etched by a 45% wt. KOH solution at 80 ºC. The substrate was left in the solution till no more bubbles from the reaction were visible around the area of the hole. A weak stream of distilled water was used to wash the sample after the etching. The chromium layer in the free-standing area was chemically removed by dipping it for a few seconds into $HClO_4$ 37%.

**Supplementary figure 3:** FWM experimental setup

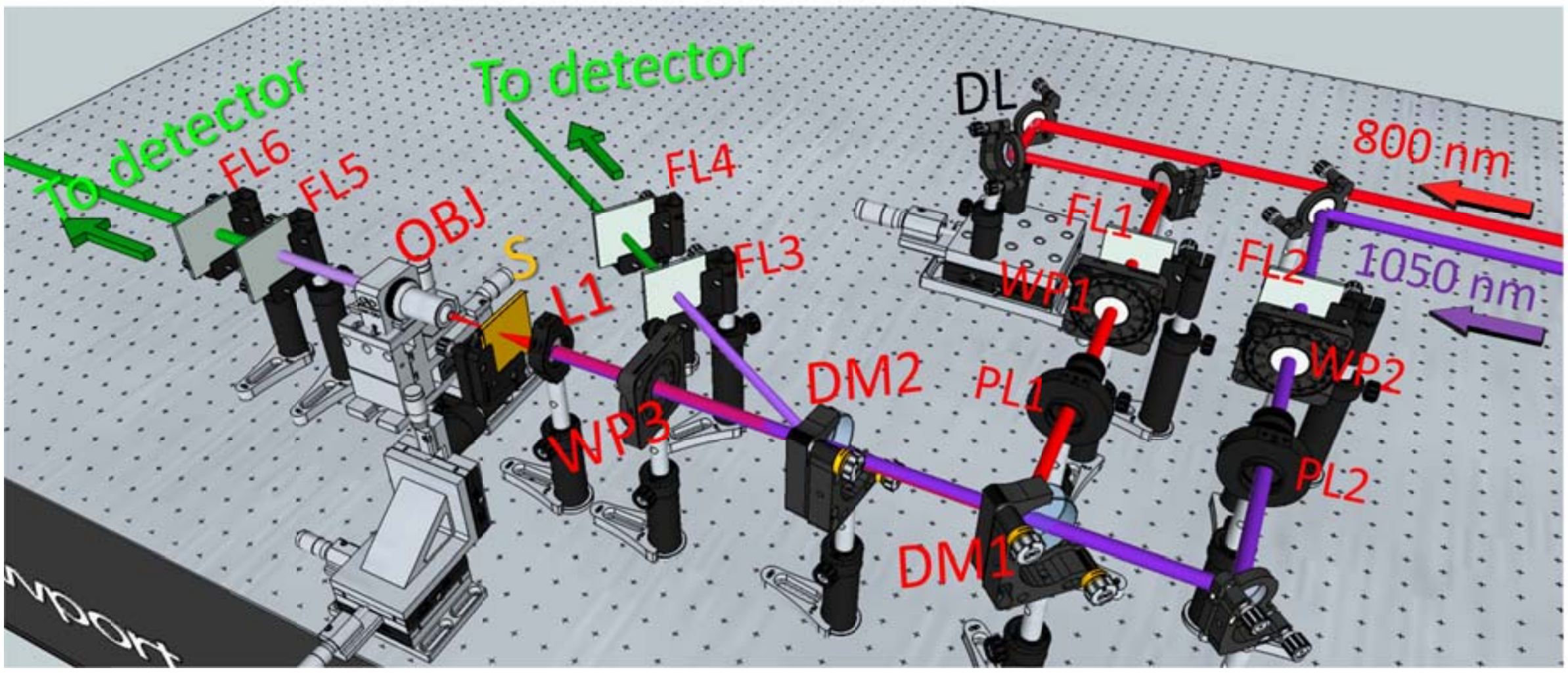

The pulses from an amplified Ti:Sapphire laser ($\omega_1$ = 800 nm, pulse duration = 60 fs, repetition rate = 1 kHz) are used to pump a tunable optical parametric amplifier (OPA) delivering pulses in the NIR with the same repetition rate and pulse duration. In our experiments, the OPA frequency was set to $\omega_2$ = 1050 nm to generate a FWM signal at $\omega_{FWM}$ = 645 nm, or to $\omega_2$ = 1265 nm to obtain FWM generation at $\omega_{FWM}$ = 585 nm. The optical path of the $\omega_1$ = 800 nm beam can be adjusted by a delay line (DL) to temporally overlap the two beams in the sample. Residual visible light in the $\omega_1$ and the OPA beams is filtered out by longpass filters (FL1 - Thorlabs FEL0750) and (FL2 - Thorlabs FEL0900). The intensity of the two beams arriving in the sample could be controlled by two identical sets of broadband half-wave plates (WP1,2 - Casix, WPZ1315) and polarizer (PL1,2 - Casix, PGL5010). The two beams are combined in the same optical path by a dichroic mirror (DM1 - Thorlabs, DMLP900). The polarization of the two beams could be jointly rotated by a half-wave plate (WP3-Casix, WPZ1315). The beams are focused on the sample (S) by a spherical lens (L1 - focal length 20 mm). The energy per pulse of the beams was set to 68 nJ at 800 nm and 12 nJ at 1050 nm. The diameter of the beams on the sample is 15 µm. The backward FWM signal is reflected at low incidence angle (with respect to the normal) by a laserline filter (DM2 - Laser components, 532NB3) positioned between the dichroic mirror and the half-wave plate, and filtered by a set of shortpass filters (FL3,4 - Thorlabs FESH0750 + Melles Griot 03SWP412). The backward FWM signal is coupled to a multimode optical fiber and sent to a CCD camera (Jobin Yvon, Symphony) connected to a spectrometer (Jobin Yvon, Triax 180). The forward FWM signal is collected by a 50X objective lens (OBJ - Mitutoyo, 378-811-3) and filtered by shortpass filters (FL5,6 - Thorlabs FESH0750 + Melles Griot 03SWP412). The signal is then coupled to a multimode optical fiber and sent to the spectrometer+CCD system.

**Supplementary discussion 1:** Nonlinear coupled mode theory simulations

The energy exchange between the longitudinal modes inside the cavities is discussed within the nonlinear coupled mode theory. The model provides a semi quantitative analysis and captures the essence of the underlying mechanism leading to the enhancement of the FWM response. The propagating waves are assumed to be stronger inside the cavity than on its surface so that most of the FWM signal is generated inside the nanoholes, as is seen by the numerical solutions of the wave equations. As the two longitudinal modes propagate inside the nanoholes, they interact with the metal and exchange energy with the generated FWM mode. The observed/measured forward FWM signal is taken as the integrated intensity of the FWM modes at the cavity exit.

The electric field of the FWM mode inside a cavity obeys the wave equation

$$\nabla^2 \boldsymbol{E}_{FWM} - \frac{\epsilon(\omega_{FWM})}{c^2}\frac{\partial^2}{\partial t^2}\boldsymbol{E}_{FWM} = \frac{1}{\varepsilon_0 c^2}\frac{\partial^2}{\partial t^2}\boldsymbol{P}^{(3)} \qquad (1)$$

where the nonlinear polarization $\boldsymbol{P}^{(3)}$ is given by eq. (1). The fields propagating inside the cavity can be expressed in terms of the dominant transverse lowest-order mode

$$\boldsymbol{E}_j(x, y, z, t) = A_0^{(j)}\boldsymbol{E}_{j,0}(x, y)e^{i(k_{j,0}z-\omega_j t)} \qquad (2)$$

with ($j$=1, 2 , FWM), $A_0^{(1)}$ and $A_0^{(2)}$ are the amplitude of the fields at $z$ = 0 and $A_0^{FWM} = A_0^{FWM}(z)$ is a slowly varying function of $z$. The propagation constants $k_{j,0} = k_{j,0}^{real} + ik_{j,0}^{img}$ and $A_0^{(j)}$ can be obtained from the 3D-FDTD calculations. As the incoming fields are polarized along the $y$ axis and hence the fields inside the cavities are mostly polarized in the same axis (according to the FDTD calculations), the tensor product in eq. 1 can be turned into scalar (the diagonal term $\chi_{yyyy}^{(3)} = \chi_d^{(3)}$ of the susceptibility tensor is much larger than the off-diagonal terms). Therefore, eqs. (2) and (1) can be plugged in eq. (1) in the main text, assuming that the FWM signal is generated mostly by Au atoms at the surfaces of the cavities and the FWM amplitude at entrance of the cavities is zero ($A_0^{FWM}(z = 0) = 0$), we get for the evolution of amplitude of the FWM mode

$$\left|A_0^{(FWM)}(z)\right| = \frac{\omega_{FWM}^2}{2c^2}\chi_d^{(3)}\left|\frac{\sin(\Delta kz/2)}{k_{FWM}\Delta k/2}\right|\left|A_0^{(1)}\right|^2\left|A_0^{(2)}\right|\Theta \tag{3}$$

where the quantity

$$\Theta = \frac{\iint_S \left|E_{1,0}(x,y)\right|^2 E_{2,0}(x,y)E_{FWM,0}(x,y)dxdy}{\iint_S \left|E_{FWM,0}(x,y)\right|^2 dxdy} \tag{4}$$

is called the spatial overlap factor and it is a degree of transverse mode matching between the longitudinal modes. In eq. (4), the integrations are performed on the cavity walls, where the fields are much stronger than anywhere else inside the metal. The phase matching condition $\Delta k = 2k_1 - k_2 - k_{FWM}$ accounts for the momentum conservation of the FWM process.

The intensity of FWM signal is given by the spatial integration of the intensity of the FWM mode at the cavity exit ($z = l$)

$$I_{FWM} \propto \iint_A |E_{FWM}(x,y,l)|^2 dxdy$$

$$= \frac{\omega_{FWM}^4 l^2}{4c^4}\left[\chi_d^{(3)}\right]^2\left|\frac{\sin(\Delta kl/2)}{k_{FWM}\Delta kl/2}\right|^2\left|A_0^{(1)}\right|^4\left|A_0^{(2)}\right|^2|\Theta|^2 e^{-2k_{FWM}^{img}l}\iint_A \left|E_{FWM,0}(x,y)\right|^2 dxdy \tag{5}$$

With the analytical approximation, the FWM process in the nanocavities is now explicitly divided into the several factors that are individually easy to understand. These factors, calculated for the different ARs using the FDTD method, are shown in the figure below. The differences in the propagation constants of the three longitudinal modes (parts a,b) causes a phase mismatch (part c) as the modes propagate inside the cavities. Combined with the attenuation of the FWM mode, it decreases the FWM efficiency. There is no explicit phase-matching resonance for a particular AR. Equally, the transverse mode profile are similar for the different AR and the spatial overlap factor (part d) weakly grows with the AR.

The analytic calculations assumed, for simplicity, that on the average the fields propagate and decay exponentially inside the nanocavities, but this is not fully supported by the numerical calculations. At the FP resonance, for example, the field decays to a minimum value and then grows again. It seems, however, that due to the phase matching, the use of an average value for the field still represents the dependence of the FWM signal on the cavity parameters.

The parameters used in the FDTD simulations used within the couple mode theory calculation are the same used in the linear FDTD simulations except for: (1) The dimensions of the mesh were set to dx = 6 nm, dy = 6 nm and dz = 3 nm for a better accuracy in the propagation direction, and (2) For the light sources we used three narrowband plane waves centered at $\omega_{FWM} =$ 645 nm, $\omega_1 =$ 800 nm and $\omega_2 =$ 1050 nm. The field amplitudes $A_0^{(1)}$ and $A_0^{(2)}$ were calculated in a plane positioned inside the holes 65 nm far from the input side. This avoids numerical artifacts arising due to sharpness of the cavities edges.

To calculate the real part $k_{j,0}^{real}$ of the propagation constant $k_{j,0} = k_{j,0}^{real} + ik_{j,0}^{img}$, two time detectors are positioned 120 nm apart inside the holes. The time spent by the pulses to travel between the detectors is calculated for the first few optical cycles only in order to avoid artifacts due to the multiple reflections inside the cavity. The imaginary part $k_{j,0}^{img}$ is obtained according to $T = (1 - R)e^{-2k_{j,0}^{img} l}$, where $T$ and $R$ are the calculated transmittance and reflectance respectively, and *l* is the cavity length (film thickness).

In conclusion, it is seen that the shape resonance at $\omega_1 =$ 800 nm, due to the fourth power dependence on the intensity of this field, is the dominant factor for the FWM field enhancement at AR = 2.

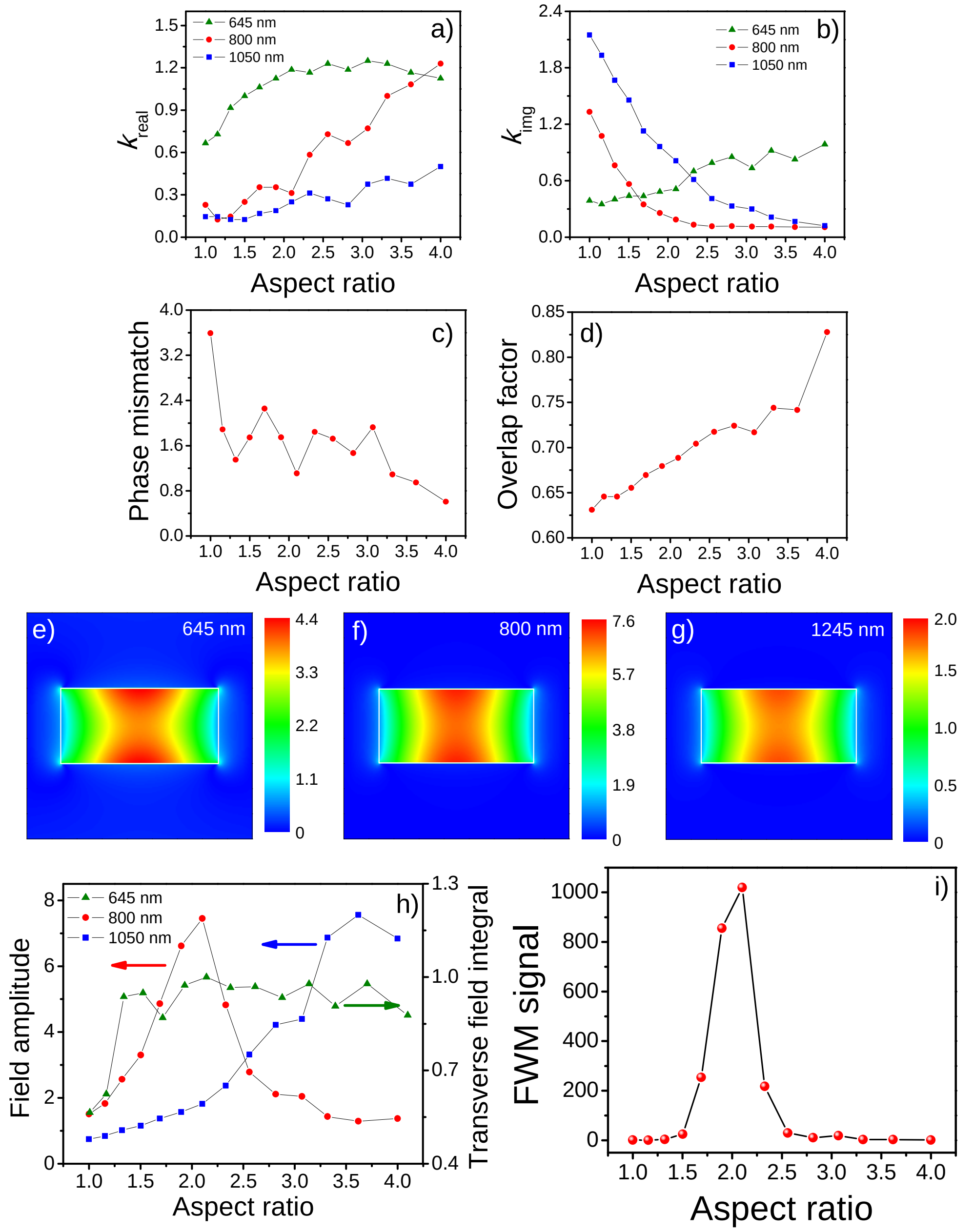

Theoretical results of the FWM shape resonance based on coupled-mode theory. Real (a) and imaginary (b) components of the cavity propagation constants $(kl)$ at $\omega_{FWM}$ = 645 nm (olive triangles), $\omega_1$ = 800 nm (red circles) and $\omega_2$ = 1050 nm (blue squares). (c) Phase mismatch factor $|sin(\Delta kl/2)/k_{FWM}(\Delta kl/2)|^2$. (d) Transverse mode overlap factor $\Theta$ in eq. 4. (e)-(g) Near-field intensity distribution at $\omega_{FWM}$ = 645 nm (e), $\omega_1$ = 800 nm (f) and $\omega_2$ = 1050 nm (g) in the entrance of a rectangle with AR = 2.1 (290x138 $nm^2$). (h) Field amplitude $A_0^{(j)}$ ($j$ = 1 – red circles, 2 – blue squares) and transverse field integral $\iint_A |E_{FWM}(x,y)|^2 dxdy$ (olive triangles). (i) Calculated FWM signal $I_{FWM}$ according to eq. (5) above.

**Supplementary Movie 1: comparison of the Fabry Perot and the LSP modes**

almeida prior movie.pptx